\documentclass[journal,comsoc]{IEEEtran}
\pdfoutput=1
\usepackage[T1]{fontenc}
\usepackage{tikz}
\usepackage{float}
\usepackage{graphicx}
\usepackage{color,soul}
\usepackage[nolist,nohyperlinks]{acronym}

\usepackage{cite}

\ifCLASSINFOpdf
\else
\fi
\usepackage{amsmath}
\begin{document}
%
\title{Relation between Functional Complexity, Scalability and Energy Efficiency in WSNs}
%
%
%

\author{Merim~Dzaferagic,
        Nicholas~Kaminski,
        Irene~Macaluso,
        and~Nicola~Marchetti\\
        CONNECT, Trinity College Dublin, Ireland, E-mail: \{dzaferam, kaminskn, macalusi, marchetn\}@tcd.ie}
\maketitle

\begin{abstract}
In order to understand the underlying mechanisms that lead to certain network properties (i.e. scalability, energy efficiency) we apply a complex systems science approach to analyze clustering in Wireless Sensor Networks (WSN). We represent different implementations of clustering in WSNs with a functional topology graph. Different characteristics of the functional topology provide insight into the relationships between system parts that result in certain properties of the whole system. Moreover, we employ a complexity metric - functional complexity ($C_F$) - to explain how local interactions give rise to the global behavior of the network. Our analysis shows that higher values of $C_F$ indicate higher scalability and lower energy efficiency.  


\end{abstract}

\begin{IEEEkeywords}
Complex systems science, functional complexity, wireless sensor networks, clustering, scalability, energy efficiency
\end{IEEEkeywords}

\IEEEpeerreviewmaketitle

\section{Introduction}\label{sec:introduction}
As \ac{WSNs} represent large deployments of unattended sensors, which are disposable and expected to last until their energy drains, energy efficiency and scalability are critical factors involved in the design of communication protocols for these devices. The physical topology is dynamic because nodes exit due to the battery discharge resulting in a constant need for adding new nodes to the network. In order to adjust to the constant changes in the physical topology, clustering was proposed in the literature as a network reorganization technique. 

Clustering partitions a network of nodes into a number of smaller groups (clusters). In addition to the energy efficiency, which explicitly affects the network lifetime, clustering algorithms have a great influence on scalability, load balancing, fault-tolerance, delay reduction, etc. \cite{Abbasi2007}. Understanding the organizational and communication characteristics of different clustering algorithms allows us to comprehend which aspects of a specific implementation lead to certain characteristics, i.e. scalability and energy efficiency. The manner in which parts of the network share information and the extent to which information spreads throughout the network represent important aspects of a clustering algorithm, as they directly impact the system longevity and the scalability of the network. The complex systems science approach we adopt allows us to investigate the amount of information about the system subparts by examining the system as a whole and comparing this to the actual amount of information that exists within the system subparts. In other words, this allows us to quantify the amount of uncertainty of interaction that exists within smaller subparts of the system compared to the uncertainty of the whole system.

Different approaches to clustering are available in the literature. The authors of \cite{Afsar2014, Aslam2012, Jiang2009, Liu2012, Nagpal98analgorithm, Ephremides1987, Xu2002, Baker1981, Lin1997, Bandyopadhyay2003, Banerjee2001, Heinzelman2002, Younis2004} introduced different approaches, which involve adaptive clustering, random competition based clustering, \ac{HCC}, energy efficient hierarchical clustering, distributed clustering, \ac{LEACH}, and \ac{HEED} clustering. As highlighted in \cite{Abbasi2007} and \cite{Younis2006}, the algorithms differ in properties like stability of the created clusters, objectives (e.g. scalability, fault-tolerance, connectivity, load balancing, redundancy elimination, rapid convergence, network lifetime), clustering criteria (e.g. identifier, position, cluster head frequency, residual energy), methodology (e.g. distributed, centralized, hybrid). We focus on the \ac{LEACH} algorithm proposed in \cite{Heinzelman2002} and the \ac{HCC} algorithm proposed in \cite{Banerjee2001}, due to the importance of these algorithms for \ac{WSNs} (the \ac{LEACH} algorithm is one of the most well-known clustering algorithms, and the \ac{HCC} algorithm is the most popular multi-tier hierarchical clustering algorithm \cite{Abbasi2007}). 

Complex systems science focuses on the underlying local interactions between system parts which give rise to the global network behavior. In \cite{Dzaferagic2016}, we propose a framework which allows us to represent network functions with graphs called functional topologies. Therein, we also propose a metric to calculate the functional complexity of an implementation of a network function. Here, we employ our functional framework to model different implementations for clustering in \ac{WSNs}. Our goal is to highlight the structural patterns present in the functional topology that result in certain properties (scalability and energy efficiency) of the implementation. This allows us to understand the underlying mechanisms that are a product of complex interactions between functional entities.

The main contributions of this paper are:
\begin{itemize}
	\item We apply a complex systems science approach to analyze clustering in \ac{WSNs};
	\item Our study shows how the density of local connections in the functional topology affects the scalability of a clustering implementation;
	\item Our results highlight the structural patterns in the functional topology that lead to higher energy efficiency of the WSN;
	\item Our functional complexity allows us to analyze the trade-off between energy efficiency and scalability of a clustering implementation.
\end{itemize}


\section{Clustering algorithms}\label{sec:Clustering algorithms}

Considering the limited energy resources and the size of \ac{WSNs}, the best approach is to partition the network into interconnected subgroups (clusters), whose local behavior gives rise to the global objective of the network (e.g. minimizing the energy consumption). Generally, clustering includes two phases:
\begin{itemize}
	\item Set-Up Phase - this phase implies choosing cluster-heads. Cluster-heads are nodes that are responsible for the coordination of the clustering process.
	\item Maintenance Phase - this phase implies reorganization of the clusters in order to add new nodes to the cluster, deal with nodes that leave the cluster, or simply change the roles of nodes (cluster-head or ordinary node) in the cluster in order to achieve higher energy efficiency \cite{Basagni1999}. 
\end{itemize}

As the maintenance phase implies the reorganization of clusters in order to achieve higher energy efficiency and to adjust to changes in the topology, we focus on this phase. As mentioned previously, we consider two clustering algorithms (the \ac{LEACH} and the \ac{HCC} algorithms) proposed in \cite{Heinzelman2002} and \cite{Banerjee2001}.

\begin{figure}[t]
	\centering
	\includegraphics[scale=0.4]{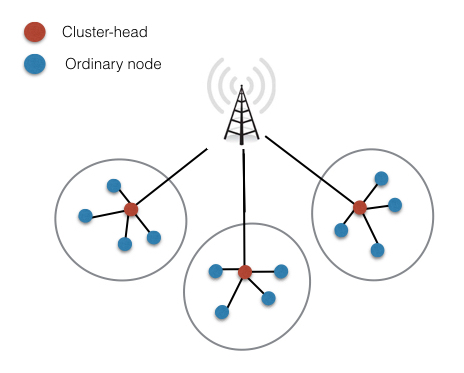}      
	\caption{The topology created according to the \ac{LEACH} algorithm. Ordinary nodes transfer sensing information to the cluster-heads, which forward this information to the base station (The Figure is redrawn from \cite{Aslam2012})}
	\label{leach_clustering_graph}
\end{figure}

The \ac{LEACH} algorithm is based on rounds. Each round includes both phases (set-up and maintenance phase). Each node in the network runs the algorithm, and decides randomly which role to play (cluster-head or ordinary node). When a node decides to be a cluster-head it broadcasts this information, whereas each node that decided to be an ordinary node listens to the broadcast messages and joins the closest cluster-head. After all ordinary nodes join one of the cluster-heads, the set-up phase finishes and the maintenance phase starts. In the maintenance phase nodes that belong to the same cluster transfer sensing information to the cluster-head (Figure \ref{leach_clustering_graph}), and as the \ac{LEACH} algorithm involves rotating the role of cluster-heads between nodes in a cluster, the nodes that belong to the same cluster communicate to each other in order to choose the next cluster-head candidate. 

\begin{figure}[t]
	\centering
	\includegraphics[scale=0.38]{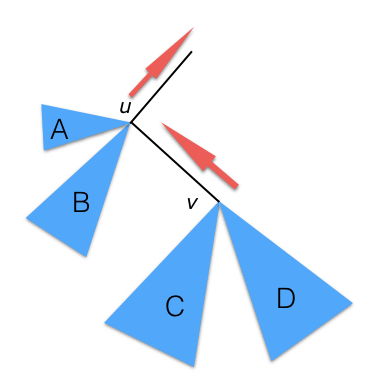}      
	\caption{Cluster formation according to the \ac{HCC} algorithm. Each node ($v, u$) discovers its subtree size and forwards this information upstream to its parent. If the subtree size is big enough, the subtree becomes a cluster (A, B, C, D). (The Figure is redrawn from \cite{Banerjee2001})}
	\label{HCC_clustering_graph}
\end{figure}

The \ac{HCC} algorithm is a multi-tier hierarchical clustering algorithm which has proven to be highly scalable. The set-up phase involves the \ac{BFS} tree discovery and cluster formation. The tree discovery is a distributed formation of a \ac{BFS} tree rooted at the initiator node. The cluster formation is shown in Figure \ref{HCC_clustering_graph}. The clustering formation process is distributed and it is executed on each node. Based on the tree discovery each node knows its parent node and its child nodes. Each node $i$ discovers the size of its downstream subtree $|V_i|$ and reports this to its parent node. The information about the subtree sizes allows us to create the clusters based on the defined cluster size $k$ (number of nodes per cluster - this number is predefined). Each node compares its subtree size $|V_i|$ to the defined cluster size $k$ and if $k \leq |V_i| < 2k$ the node initiates the cluster formation on its subtree. As the \ac{BFS} tree denotes the routes to the base station, the sensing information of all sensors is transmitted using these routes. The maintenance phase involves just the maintaining up to date information on each node about its parent and child nodes. 

\section{Functional framework}\label{sec:Functional framework}
In order to analyze the underlying connectivity patterns of a clustering algorithm, we employ the functional framework introduced in \cite{Dzaferagic2016}. The framework allows us to represent the implementation of a clustering algorithm with a graph called functional topology. We create functional topologies based on the functional connectivity between system parts, where each node represents a functional entity or any information source related to the implementation of the particular clustering algorithm, and each link indicates interactions between nodes. To quantify the variety of connection patterns between system entities and the roles that these entities have in the topology we employ a metric called functional complexity \cite{Dzaferagic2016}. The multi-scale functional complexity is calculated with equation (\ref{eq:complexity}).

\begin{equation}\label{eq:complexity}
C_F = \dfrac{1}{R-1}\displaystyle\sum_{r =1}^{R-1}\sum_{j = 1+r}^{N} | \langle I_r(\Lambda^j) \rangle - \dfrac{j}{N}  I_r(\Lambda^N)|
\end{equation}

\begin{table}[t]
	\centering
	\caption{}
	\label{tab:equation_notations}
	\begin{tabular}{c|p{\dimexpr 0.6\linewidth-2\tabcolsep}}
		\textbf{Symbol} & \textbf{Meaning}    \\ \cline{1-2}
		\centering{$N$} & total number of nodes in the functional topology
		\\ \cline{1-2}
		\centering{$j$} & subgraph size - number of nodes in the subgraph
		\\ \cline{1-2}
		\centering{$r$} & scale size
		\\ \cline{1-2}
		\centering{$R$} & maximum scale size, which is defined as the
		longest shortest path in the whole functional topology
		\\ \cline{1-2}
		\centering{$H(x_n)$} & entropy of node $n$ which indicates the
		uncertainty of interactions of node $n$ in the operation of a
		network function
		\\ \cline{1-2}
		\centering{$\Lambda_k^j$} & $\mathrm{k^{th}}$ subgraph with $j$ nodes 
		\\ \cline{1-2}
		\centering{$I_r(\Lambda^N)$} & the total amount of information of the subgraph with $N$ nodes for scale $r$
		\\ \cline{1-2}
		\centering{$\langle  I_r(\Lambda^j) \rangle$} & the average amount of information over all subgraphs with the size $j$
	\end{tabular}
\end{table}

The meaning of the terms appearing in equation (\ref{eq:complexity}) is shown in Table \ref{tab:equation_notations}. As \ac{WSNs} rely on the communication established between one hop neighbors, in our analysis the maximum scale size will be equal to $R = 2$. Hence, we perform a single-scale analysis with a simplified version of equation (\ref{eq:complexity}), which is represented with equation (\ref{eq:single_scale_complexity}). 

\begin{equation}\label{eq:single_scale_complexity}
C_F = \displaystyle\sum_{j = 2}^{N} | \langle I(\Lambda^j) \rangle - \dfrac{j}{N}  I(\Lambda^N)|
\end{equation}

$I_r(\Lambda_k^j)$ is the total amount of information of the  $\mathrm{k^{th}}$ subgraph with $j$ nodes. It is calculated with equation (\ref{eq:total_amount_of_information}).

\begin{equation}\label{eq:total_amount_of_information}
I_r(\Lambda_k^j) = \displaystyle\sum_{n \in \Lambda_k^j} H(x_n)
\end{equation}

The functional complexity compares the uncertainty of interactions for a smaller subset ($\langle I_r(\Lambda^j) \rangle$) to the uncertainty which is expected from the calculation performed on the whole system ($I_r(\Lambda^N)$). $H(x_n)$ reaches its maximum if the probability of interaction with node $n$ is $p(x_n = 1) = 1/2$. As the distribution of links among nodes for a sparse graph is almost uniform, a sparse graph results in high values of $H(x_n)$. High values of $H(x_n)$ result in high values of $I_r(\Lambda_k^j)$. Therefore, the functional complexity is high for a sparse graph, with uniformly distributed links among nodes for subgraphs with the size $j < N$. The functional complexity is zero for a fully connected and for a disconnected graph. For more details about the functional framework and the complexity metric expressed by equation (\ref{eq:complexity}) the reader is referred to \cite{Dzaferagic2016}. 

\begin{figure}[t]
	\centering
	\includegraphics[scale=0.45]{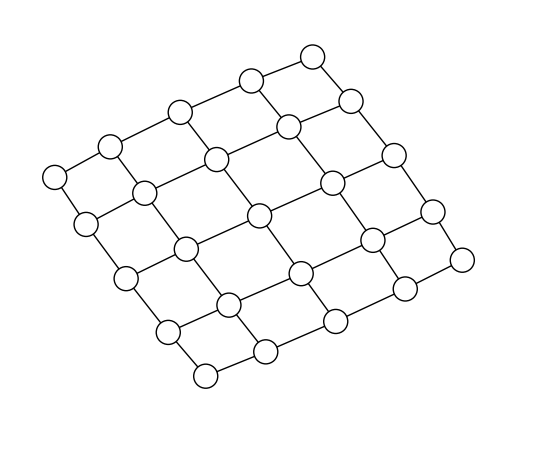}      
	\caption{An example of a physical topology of a \ac{WSN} according to the Von Neumann neighborhood.}
	\label{physical_topology}
\end{figure}

The graph shown in Figure \ref{physical_topology} is an undirected graph created according to the Von Neumann neighborhood, and it depicts an example of a physical topology of a \ac{WSN}. A wireless node $A$ can transmit information to another wireless node $B$ only if node $B$ is within the transmission radius $R_A$ of node $A$. In the case of \ac{WSNs}, we consider the communication between two nodes established only if both nodes can transmit information to each other. In other words, we consider that two nodes can exchange information only if the distance between them is $d(A,B)  \leq Min\{R_A, R_B\}$.  

In \cite{Dzaferagic2016}, we presented an approach to map different frequency allocation algorithms into functional topologies. We apply the same approach to examine the functional topologies of the \ac{LEACH} and \ac{HCC} algorithms. 

\begin{figure}[t]
	\centering
	\includegraphics[scale=0.5]{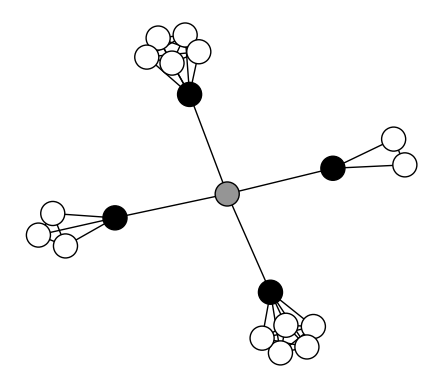}      
	\caption{The functional topology of the \ac{LEACH} algorithm for a network of twenty nodes which are divided into four clusters. The white nodes represent ordinary nodes, the black nodes represent cluster-heads, and the gray node represents the base station. }
	\label{functional_topology_leach}
\end{figure}

Our goal is to investigate the influence of the interactions among nodes after the clusters are established, on the objectives of clustering algorithms (specifically scalability and energy efficiency). We start with the \ac{LEACH} algorithm. As discussed before, during the maintenance phase of the \ac{LEACH} algorithm ordinary nodes that belong to the same cluster send their sensing information to the cluster-head, and they talk to each other in order to decide who is the next candidate for the cluster-head role. According to the approach in \cite{Dzaferagic2016}, we imagine a virtual decision maker entity that is moving from one node to another. At each ordinary node the decision maker entity communicates with the cluster-head and all other ordinary nodes that belongs to the same cluster. At each cluster-head the decision maker entity communicates with each ordinary node that belongs to the same cluster and with the base station. This results in a functional topology with dense local connections (Figure \ref{functional_topology_leach}). 

\begin{figure}[t]
	\centering
	\includegraphics[scale=0.55]{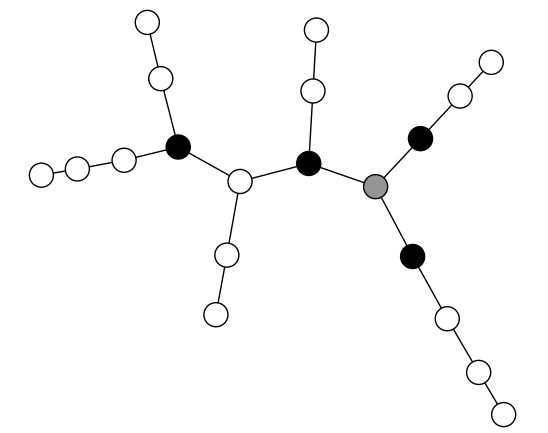}      
	\caption{The functional topology of the \ac{HCC} algorithm for a network of twenty nodes, which represents the \ac{BFS} tree created based on the physical topology shown in Figure \ref{physical_topology}. The white nodes represent ordinary nodes, the black nodes represent cluster-heads, and the gray node represents the base station. }
	\label{functional_topology_HCCA}
\end{figure}

To examine the functional topology of the \ac{HCC} algorithm we use the same approach presented in the previous example. Again we focus on the maintenance phase of the algorithm. After the set-up phase the nodes establish connections to their neighboring nodes according to the \ac{BFS} algorithm. Each node discovers its subtree, and exchanges information with its neighbors in the \ac{BFS} tree. Again, we imagine a virtual decision maker entity that is moving from one node to another. At each node the decision maker entity collects information from nodes that belong to its subtree  and forwards this information to its parent. In other words, each node maintains its position in the \ac{BFS} tree and therefore the functional topology of the \ac{HCC} algorithm is equivalent to the \ac{BFS} tree of the physical topology (Figure \ref{functional_topology_HCCA}).


\section{Analysis}\label{sec:Analysis}
In order to transfer the information collected at a sensor node to a base station, \ac{WSNs} establish paths in an ad-hoc manner. Two nodes can exchange information if and only if they are within the transmission radius of each other. However, after the set-up phase of the clustering algorithm finishes, nodes do not communicate to all other nodes that are within their coverage area. At this stage, the information exchange depends on the rules of the clustering algorithm. 

The topology of \ac{WSNs} is subject to constant changes due to the disposable nature of wireless sensors and the constant need to expand and densify the sensing area. Therefore, besides the energy efficiency, clustering algorithms for \ac{WSNs} have to exhibit scalability. Scalability is the capability of the network to adapt to new nodes joining the network, existing nodes leaving the network, and other nodes migrating from one cluster to another \cite{Banerjee2001}. 

The maintenance phase of the \ac{LEACH} algorithm invokes the set-up phase, where nodes that belong to the same cluster talk to each other to decide which node is the next candidate for the cluster-head role, and each node transfers sensing information to the current cluster-head. Adding a new node to any cluster means establishing a connection to all nodes within the cluster because each node has to be aware of all other nodes that belong to the same cluster, in order to maintain the process of cluster-head elections. 

The maintenance phase of the \ac{HCC} algorithm is simpler than that of the \ac{LEACH} algorithm. Each node forwards the information of its children and transfers its own sensing information to its parent node. Therefore, adding a new node to the cluster simply means that the new node gets connected to one of the existing nodes which is going to be its parent node. The new node does not need to inform all nodes in the cluster about its arrival, which simplifies the process of adding nodes to the network.

The authors of \cite{Afsar2014,Aslam2012,Jiang2009,Liu2012} and \cite{Pantazis2013} discuss different objectives of clustering algorithms. Among other things, the authors focus on scalability issues and network longevity (energy efficiency) of these algorithms. Hierarchical approaches proved themselves to be more scalable than their non-hierarchical counterpart. The authors of \cite{Jiang2009} showed that a trade-off exists between network scalability and energy consumption in clustering schemes. Our goal is to investigate the relationship between these objectives and the functional complexity, which would allow us to analyze them together. In \cite{Afsar2014,Aslam2012,Jiang2009,Liu2012} and \cite{Pantazis2013}, the authors emphasize that the \ac{LEACH} algorithm is designed to extend the network longevity, whereas the \ac{HCC} algorithm targets scalability as the main objective. In other words, the \ac{LEACH} algorithm is less scalable and the \ac{HCC} algorithm is less energy efficient. 

As the authors of \cite{Abbasi2007,Afsar2014,Aslam2012,Jiang2009,Liu2012} agree that the communication between the cluster-head and the base station consumes most energy, we propose to calculate the energy efficiency of a clustering implementation as the ratio between the average number of intra-cluster connections and the number of links between the base station and each cluster-head in the functional topology. If the ratio increases the energy efficiency increases, due to the bigger number of intra-cluster connections as compared to the number of connections between the base station and the cluster-heads. As the scalability of the networks represents the adaptability to changes, we calculate it as the average number of messages sent when a new node joins the network. If the average number of messages increases the scalability of the network decreases. 

\begin{figure}[t]
	\centering
	\includegraphics[scale=0.4]{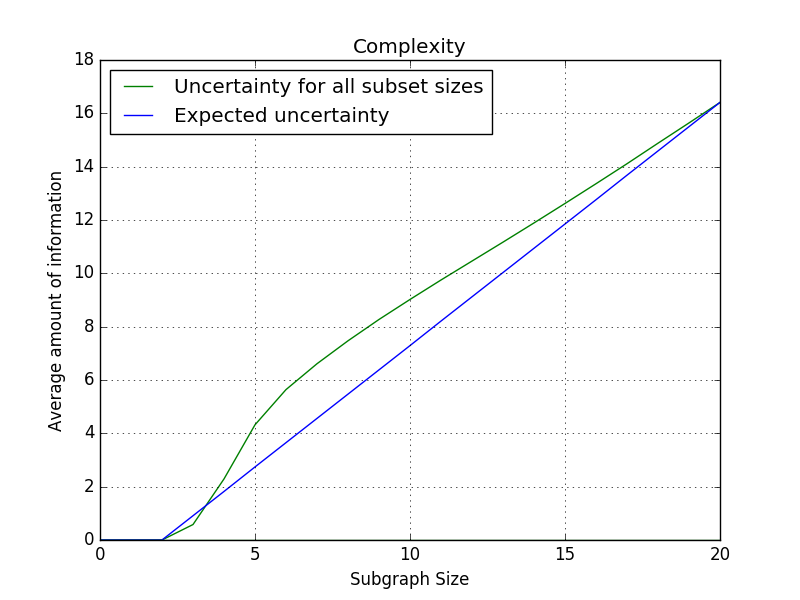}      
	\caption{Functional complexity of the \ac{LEACH} algorithm; the functional topology has twenty nodes and the nodes are divided into 4 clusters; the maximum scale size R is 2; the functional complexity is the area between the green and the blue curves.}
	\label{complexity_leach_4x5}
\end{figure}

Figure \ref{complexity_leach_4x5} depicts the relationship between the uncertainty of interactions for all subset sizes and the uncertainty which is expected from the calculation performed on the whole system for the \ac{LEACH} algorithm, i.e. the functional complexity of LEACH. As shown in Figure \ref{functional_topology_leach} the functional topology of the \ac{LEACH} algorithm has dense local connections (intra-cluster connections), whereas the inter-cluster connections are sparse. The dense intra-cluster connections result in low scalability of the algorithm, due to the need of interacting with all nodes that belong to the cluster in order to add a new node. In other words, in order to add a node to a cluster, all nodes that belong to the cluster have to update their information. Our analysis shows that for a network of twenty nodes which are divided into four clusters the average number of messages sent when a node joins the network is 3.75. The dense local connections result in smaller values of the average uncertainty of interactions for subsets with size three than we expected from the calculation performed on the whole topology. Figure \ref{complexity_leach_4x5} shows that the uncertainty for subgraphs with the size greater than or equal to four is much higher than expected from the uncertainty of the whole system. This is because of the sparse inter-cluster connections, due to which we can find a lot of subgraphs with these sizes which have uniform distributions of links among nodes. For example, in Figure \ref{functional_topology_leach}, if we choose any two ordinary nodes that belong to different clusters, we can create a sparse subgraph with size five. The functional complexity of the \ac{LEACH} algorithm is 19.24, which is, as we will see below, relatively low compared to an algorithm that is highly scalable. 

\begin{figure}[t]
	\centering
	\includegraphics[scale=0.4]{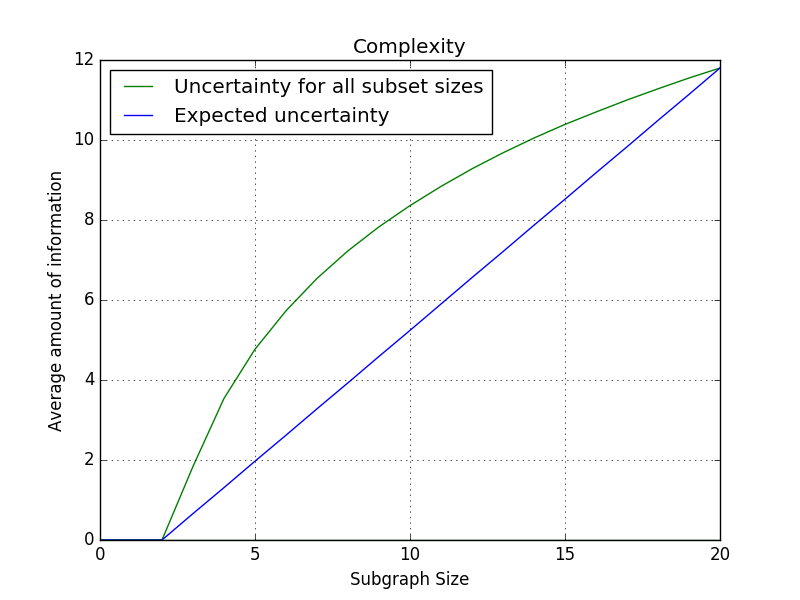}      
	\caption{Functional complexity of the \ac{HCC} algorithm; the functional topology has twenty nodes; the maximum scale size R is 2; the functional complexity is the area between the green and the blue curves.}
	\label{complexity_hcc_4x5}
\end{figure}

Figure \ref{complexity_hcc_4x5} depicts the functional complexity of the \ac{HCC} algorithm. As shown in Figure \ref{functional_topology_HCCA} the functional topology of the \ac{HCC} algorithm has sparse intra and inter cluster connections. As the links in the functional topology represent functional dependencies between nodes, a sparse connectivity pattern indicates weak dependencies which result in high scalability. This follows from the fact that in order to add a node to the network, the new node needs to establish a connection (send a message) to one of the nodes in the topology and to declare this node as its parent. Therefore, the information about a new node joining the network does not have to be transmitted throughout the cluster. The functional complexity of the \ac{HCC} algorithm is 38.31, which is high compared to the functional complexity of the \ac{LEACH} algorithm. The energy efficiency of the \ac{HCC} algorithm is 0.61, which is low compared to the \ac{LEACH} algorithm (energy efficiency is 1.08) for the same number of nodes which are divided into four clusters. 

Table \ref{tab:leach_table_complexities} shows how the functional complexity, the energy efficiency and the scalability of the \ac{LEACH} algorithm change for different number of clusters. As shown the functional complexity increases until the number of clusters reaches sixteen, and then it starts decreasing. This is due to the fact that if the number of nodes is fixed, increasing the number of clusters decreases the number of nodes per cluster, which makes the graph more and more sparse until it reaches a point where the graph is not clustered any more and each node talks directly to the base station, i.e. a star graph. If the number of nodes per cluster decreases, the number of intra-cluster connections decreases, and hence, the implementation becomes more scalable. On the other hand, the implementation becomes less energy efficient, due to the increasing number of nodes that communicate directly to the base station. 

Changing the number of clusters for the \ac{HCC} algorithm does not affect the functional complexity of the implementation, because the functional topology does not change. As each node transfers information to its parent node, increasing the number of clusters decreases the number of nodes per cluster, but the information paths do not change (and therefore the functional topology does not change). As the information paths do not change and the information travels according to the established \ac{BFS} routes, the main purpose of clustering is to logically group nodes in order to provide scalability (due to the hierarchical control architecture) and enable aggregation/fusion of the sensing information at the chosen cluster-heads. 

\begin{table}[t]
	\centering
	\caption{Functional complexity, scalability and energy efficiency of the \ac{LEACH} algorithm for different number of clusters; the functional topology has twenty nodes.}
	\label{tab:leach_table_complexities}
	\begin{tabular}{l|l|l|l|l|l|l}
		\#of clusters    & 3     & 4     & 5     & 6     & 16 & 19    \\ \hline
		C\_F              & 14.35 & 19.24 & 22.69 & 25.55 & 32.4 & 31.85 \\ \hline
		Energy efficiency & 1.91  & 1.08  & 0.72  & 0.51  & 0.07 & 0.05  \\ \hline
		Avg. \#msg. if node joins & 5.33  & 3.75  & 2.8   & 2.16  & 1 & 1    
	\end{tabular}
\end{table}

Our analysis confirms the observation made by the authors of \cite{Jiang2009}, which highlights the trade-off between scalability and energy efficiency. With our complex systems science approach we showed that these aspects of different clustering algorithms can be analyzed together by analyzing the functional complexity of the specific implementation. According to Table \ref{tab:leach_table_complexities}, increasing values of $C_F$ lead to the increase of scalability and the decrease of energy efficiency.

\section{Conclusion}\label{sec:conclulsion}
The growing interest in \ac{WSNs} due to the various applications in IoT results in a great variety of clustering algorithms that support these applications. Very often, one compares these algorithms based on certain properties (e.g. fault-tolerance, delivery delay, energy efficiency, scalability). Comparing these algorithms based on a broad range of properties is very difficult. Additionally, a lack of understanding exists when we try to comprehend the mechanisms that lead to these properties. The aim of this paper is to apply a complex systems science approach to analyze these mechanisms. We use the functional framework that we proposed in \cite{Dzaferagic2016} to model different implementations of clustering in \ac{WSNs}. The framework allows us to represent the implementations of network functions with graphs, which are called functional topologies.

The next step after mapping the implementations of clustering into functional topologies is to calculate the functional complexity ($C_F$) of these topologies. The functional complexity captures the variety of structural patterns in the topology and quantifies the deviation in uncertainty of interactions for a smaller subset of the system from the uncertainty which is expected from the calculation performed on the whole system. In other words, $C_F$ quantifies how much information we can not capture simply by studying the whole system, due to the complex relationships that exist between smaller subsets of the system. 

In this paper we focus on two clustering algorithms, i.e. the \ac{LEACH} and the \ac{HCC} algorithms. Our goal is to investigate the mechanisms that provide high scalability in the case of the \ac{HCC} algorithm and high energy efficiency for the \ac{LEACH} algorithm. We also study the trade-off between these two network properties. After briefly introducing the algorithms, we map their implementations into functional topologies, and calculate the corresponding functional complexities.

We show that high functional complexity indicates greater scalability of the implementation. We also show that increasing values of the functional complexity with the number of clusters, indicates lower energy efficiency. We then highlight that a sparse functional topology results in higher complexity. Therefore, our functional topology explicitly explains both the higher scalability (sparse graph) and lower energy efficiency of the \ac{HCC} algorithm as compared to the \ac{LEACH} algorithm.

\section*{Acknowledgment}
This material is based upon works supported by the Science Foundation Ireland under the Grant No. 13/RC/2077.

\ifCLASSOPTIONcaptionsoff
  \newpage
\fi


\bibliographystyle{templates/IEEEtran}  
\bibliography{clustering_in_wsn}



\begin{acronym}
	\acro{RNC}{Radio Network Controler}
	\acro{nodeB}{UMTS base transciever station}
	\acro{VLR}{Visitor Location Register} 
	\acro{HLR}{Home Location Register}
	\acro{SGSN}{Serving GPRS Support Node}
	\acro{MSC}{Mobile Switching Center}
	\acro{SMS}{Short Message Service}
	\acro{IoT}{Internet of Things}
	\acro{AP}{Access Points}
	\acro{WSNs}{Wireless Sensor Networks}
	\acro{WSN}{Wireless Sensor Network}
	\acro{IoT-GSI}{Global Standards Initiative on Internet of Things}
	\acro{DCA}{Distributed Clustering Algorithm}
	\acro{BFS}{Breadth-First Search}
	\acro{LEACH}{Low Energy Adaptive Clustering Hierarchy}
	\acro{HCC}{Hierarchical Control Clustering}
	\acro{HEED}{Hybrid Energy-Efficient Distributed}
\end{acronym}

\end{document}